\documentclass[%
twocolumn,
 reprint,
superscriptaddress,
 amsmath,amssymb,
 aps,
prd,
longbibliography,
]{revtex4-1}

\usepackage{aas_macros}
\usepackage{graphicx}
\usepackage{dcolumn}
\usepackage{bm}
\usepackage{hyperref}
\usepackage{orcidlink}

\usepackage{color}
\usepackage[mathlines]{lineno}
\usepackage{tabularx}
\usepackage{graphicx}
\usepackage{subfigure}
\usepackage[normalem]{ulem}
\usepackage[export]{adjustbox} 
\usepackage{url}
\usepackage{subcaption}
\DeclareUnicodeCharacter{2009}{\,} 

\hypersetup{
    pdfnewwindow=true,    
    colorlinks=true,      
    linkcolor=blue,       
    citecolor=blue,       
    filecolor=blue,       
    urlcolor=blue         
}

\usepackage{natbib}

\begin{document}
\title{BEC vortices as an observational signature of ultra-light bosonic dark matter}
\author{Rongzi Zhou,\orcidlink{0000-0002-4222-6185}}
\email{rzzhou@link.cuhk.edu.hk}
\affiliation{Department of Physics, The Chinese University of Hong Kong, Shatin, New Territories, Hong Kong, China}
\author{Dylan M. H. Leung, \orcidlink{0009-0007-5539-2151}}
\thanks{Now at Ohio State University: \href{mailto:leung.253@buckeyemail.osu.edu}{leung.253@buckeyemail.osu.edu}}
\affiliation{Department of Physics, The Chinese University of Hong Kong, Shatin, New Territories, Hong Kong, China}
\affiliation{Center for Cosmology and AstroParticle Physics (CCAPP), Ohio State University, Columbus, OH 43210}
\affiliation{Department of Physics, Ohio State University, Columbus, OH 43210}
\author{Jason S. C. Poon}
\email{jason.poon@link.cuhk.edu.hk}
\author{Ming-Chung Chu}
\email{mcchu@phy.cuhk.edu.hk}
\affiliation{Department of Physics, The Chinese University of Hong Kong, Shatin, New Territories, Hong Kong, China}

\date{December 1, 2024}

\begin{abstract}

Ultra-light bosonic dark matter (ULDM) is an interesting and promising dark matter candidate. While the wave-like nature of ULDM has been widely studied in the literature, we explore another distinctive feature of ULDM as Bose-Einstein Condensate (BEC) in this paper: the emergence of vortices in a rotating BEC-ULDM halos. Using numerical solution of the GPP equation, we demonstrate that a lattice of vortices — underdensity columns that carry angular momentum — naturally forms in a ULDM halo under conditions similar to those of the Milky Way. Furthermore, we study the gravitational lensing by these vortices as a possible observational signature of BEC-ULDM. If the vortices are large enough and the halo's rotational axis align with the line of sight, regularly separated brightness anomalies can be produced, providing strong evidence for BEC-ULDM.

\end{abstract}

\maketitle

\section{Introduction} \label{sec:Introduction}
\par
The existence of dark matter (DM) in the universe is strongly supported by extensive astronomical evidences, including galaxy rotation curves and gravitational lensing effects. Despite this, the properties of dark matter remain unknown. A wide range of candidates has been proposed, spanning from ultra-light particles with masses of order $10^{-22}~\text{eV}$ to weakly interacting massive particles (WIMPs) with masses of tens of GeV or more, and even solar mass primordial black holes.


The traditional cold dark matter (CDM) model, which assumes DM to be collisionless classical particles, has been successful in predicting large-scale cosmic structures consistent with astrophysical observations~\cite{Spergel:1999mh, BOSS:2013rlg}. However, it faces significant challenges on smaller scales, such as the cusp-core problem~\cite{de2010core, genina2018core}. These discrepancies have motivated researchers to explore alternative models, including the intriguing possibility of ultralight bosonic dark matter (ULDM).

Due to its extremely small particle mass $m\approx10^{-22}~\text{eV}$, ULDM exhibits wave-like properties on galactic scales, as the associated de Broglie wavelength can extend to about a kiloparsec:

\begin{equation}
    \lambda_{\text{db}}(r) = 0.48~\text{kpc} \left(\frac{10^{-22}~\text{eV}}{m}\right)\left(\frac{250~\text{km/s}}{v}\right),
\end{equation}

where $v$ is the particle's speed relative to the galactic rest frame. This wavelength, which exceeds the typical inter-particle separation in a DM halo, suggests that galactic-sized Bose-Einstein condensates (BECs) could form in the central region of a DM halo~\cite{Hui:2016ltb, Boehmer:2007um, Schive:2014hza}. Such structures naturally suppress small-scale density fluctuations, producing a cored density profile in the galaxy center. Furthermore, cosmological simulations of ULDM demonstrate its ability to reproduce the observed large-scale structure of the universe, comparable to the predictions of the CDM model~\cite{Matos:2000ng, Sahni:1999qe, Schive:2014dra}.

The BEC core at the center of a ULDM halo exhibits superfluidity. In this context, the velocity is defined as the gradient of the phase, $\boldsymbol{v} = \nabla S$, of the wave function. If $\oint \boldsymbol{v} \cdot dl = 2 j \pi$ (equivalently, a non-vanishing curl, $\vec{\nabla} \times \boldsymbol{v}$), with $j$ an non-zero integer, vortices would be formed. These vortices will result in localized structures where the density drops to zero due to destructive interference. It is natural for a DM halo to acquire non-zero angular momentum from tidal torque. Consequently, ULDM halos are expected to form vortices once their angular momenta reach a critical value.

The formation of vortex structures in BECs is well-studied in laboratory experiments, typically on scales of $50$--$100~\mu\text{m}$, where multiple vortices arrange themselves into rings or lattices~\cite{2006Prama..66....3S, fetter2001vortices, schweikhard2004vortex}. This unique feature of ULDM—the ability to form vortices—distinguishes it from other dark matter candidates and could serve as a key observational signature. To this end, we propose gravitational lensing as a potential method for observing such vortices. Gravitational lensing, determined entirely by the mass distribution of the lensing object (e.g., a galaxy or galaxy cluster), is a powerful tool for probing dark matter structures. Moreover, it is sensitive to not only the macrostructure of the lens galaxy but also local substructures near the lensed images~\cite{10.1046/j.1365-8711.1998.01319.x, refId0}. Therefore, vortex structures may alter the fluxes of lensed images in ways that cannot be explained by smooth lens profiles without substructures, akin to the well-known flux ratio anomaly~\cite{10.1046/j.1365-8711.1998.01319.x}. Moreover, vortices are underdensity features, which would lead to different lensing effects as those of the overdense subhalos expected in standard CDM scenarios. 

In this paper, we focus on simulating the emergence of vortices in ULDM halos and investigating their potential observational significance through gravitational lensing. In Section~\ref{sec:Formulation}, we review the formalism for calculating the ULDM core and the basics of gravitational lensing. In Section~\ref{sec:Method}, we outline the preconditioned conjugate gradient method used for vortex simulations, along with details of the simulation setup. Section~\ref{sec:Result} presents our key results, including vortex formation and the corresponding gravitational lensing effects. Finally, in Section~\ref{sec:Conclusions}, we summarize our findings and discuss potential directions for future research.

\section{Formulation} 
\label{sec:Formulation}
Under the mean field approximation, a system of bosons in BEC can be described by a single macroscopic wavefunction:

\begin{equation}
    \Psi(r,t) = \sqrt{\rho(r,t)} e^{i S (r,t)/ \hbar}\,,
\end{equation}
where the amplitude of the wavefunction determines the number density profile $\rho(r,t)$, and $S(r,t)$ is the phase. The wavefunction satisfies the normalization condition:

\begin{equation}
    \int |\Psi|^2 \, d^3r = N \,,
\end{equation}
where $N$ is the number of particles.

\subsection{Gross-Pitaevskii-Poisson Equation}

The virial speed for a DM halo is much lower than the speed of light, allowing us to study the system in the non-relativistic regime. Consequently, the Klein-Gordon equation reduces to the Gross-Pitaevskii Equation (GPE), together with the gravitational Poisson equation:

\begin{align} \label{eq:GPP}
    ih \frac{\partial \Psi}{\partial t} &= \left[\frac{-\hbar^2}{2m}\nabla^2 + m(\Phi + \Phi_{\text{Bary}}) + g|\Psi|^2 - \Omega L_z\right] \Psi \,,
\end{align}
\begin{align}  \label{eq:Poisson}
    \nabla^2\Phi &= 4\pi Gm|\Psi|^2 \,.
\end{align}


Eq.~\ref{eq:GPP} is a non-linear Schrödinger equation that describes the dynamics of the macroscopic wavefunction $\Psi$, representing the system of ULDM bosons at very low temperatures. The first term on the right hand side represents the kinetic energy acting on the system.
The second term corresponds to the gravitational energy, where $\Phi$ ($\Phi_{\text{Bary}}$) is the gravitational potential of DM (baryonic matter). The third term is the self-interaction energy, which is chosen to be repulsive in this study, and its strength is proportional to the scattering length $a_s$:
\begin{equation}
    g = \frac{4\pi \hbar^2 a_s}{m} > 0 \,.
\end{equation}
The fourth term describes the rotational energy of the system, with $\Omega$ representing the angular speed and $L_z$ the angular momentum along the rotation axis, taken to be the $z$ direction.

Eq.~\ref{eq:Poisson} is the Poisson equation, which links the gravitational potential $\Phi$ to the density $m|\Psi|^2$, creating another source of non-linearity in the GPE.

Since the system is scale-free, Eq.s~\ref{eq:GPP} and~\ref{eq:Poisson} are invariant under these scaling:

\begin{eqnarray} \label{eq:scale}
    t \rightarrow \lambda^2 t \,, \quad r \rightarrow \lambda r \,, \quad g \rightarrow \lambda^{-2}\kappa^{-2} g \,, \\[1.5em]
    G \rightarrow \lambda^{-4}\kappa^{-2} G \,, \quad \Psi \rightarrow \kappa \Psi\, ,
\end{eqnarray}
where the parameters $\kappa$ and $\lambda$ serve as scaling factors.

\begin{figure}[t!]
    \centering
    \includegraphics[width=83mm]{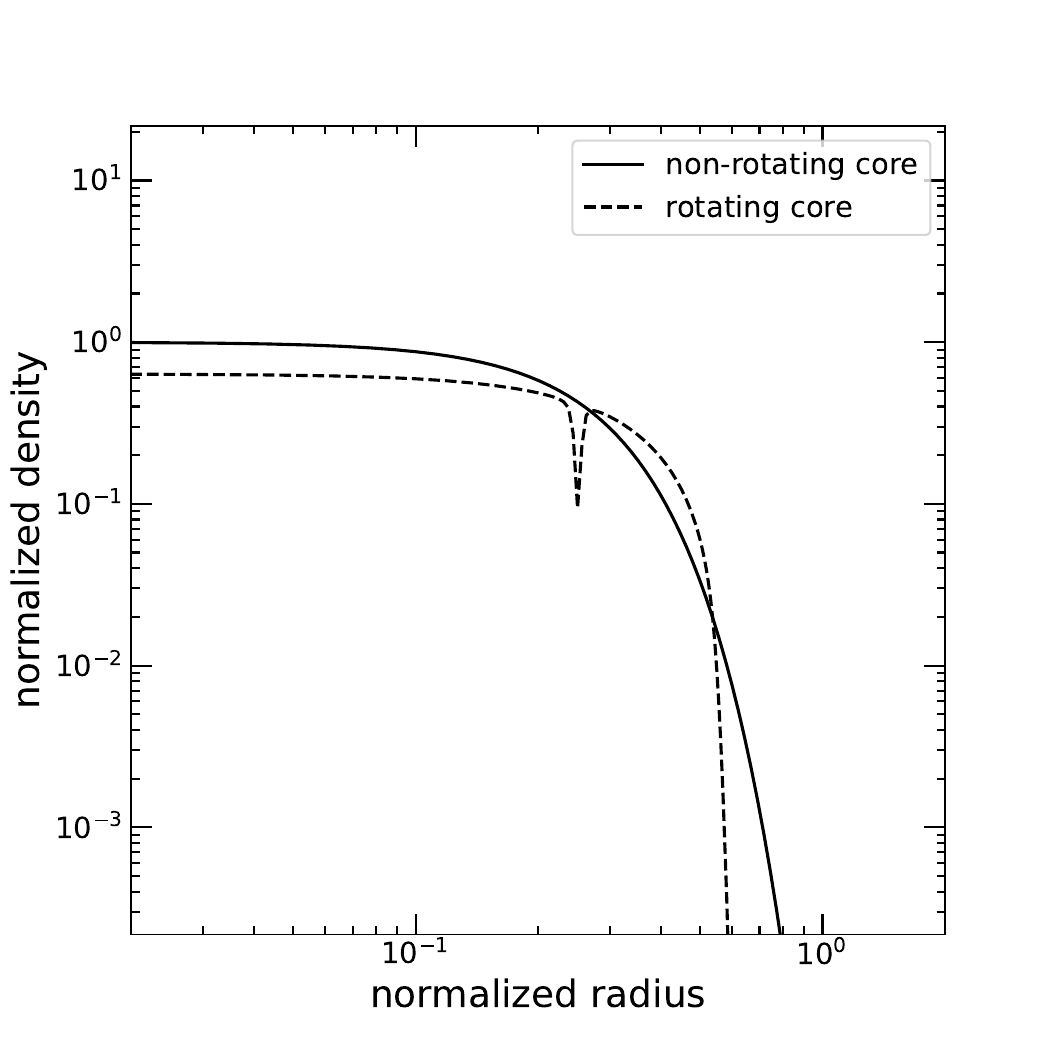}
    \caption{Comparison of the soliton density profile with (dashed line) and without(solid line) rotation. For the rotating case, we show the normalized density profile along the $x$ axis. The non-rotating density profile can be described by the TF approximation from Eq.~\ref{eq:TF}. A vortex shows up as a local under-density region in the rotating soliton.}
    \label{fig:TF}
\end{figure}

\subsection{Analytical Solution}

Under the Thomas-Fermi (TF) approximation, where the interaction energy is much larger than the kinetic energy for a large number of particles, the kinetic energy term in Eq.~\ref{eq:GPP} can be neglected. If we further assume that $\Omega$ is zero for simplicity, the TF solution can be obtained as follows~\cite{Rindler-Daller:2011afd}:

\begin{equation}\label{eq:TF}
    \rho_{\text{TF}}(r) =  \rho_c \, \frac{\sin\left(\dfrac{r}{R_0}\right)}{\left( \dfrac{r}{R_0}\right)}\,,
\end{equation}
where $\rho_c$ is the central density, and

\begin{equation}
    R_0 = \pi \sqrt{\dfrac{g}{4\pi G m^2}}\,.
\end{equation}
The TF approximation is not the exact solution for the realistic rotating BEC soliton, but it is a good initial guess for the simulation.

Fig.~\ref{fig:TF} shows the non-rotating pure ULDM core density profile predicted under the TF approximation, which is more consistent with observational results than the cuspy profile predicted by the CDM model. Fig.~\ref{fig:TF} also shows the rotating case where a vortex appears as a local underdensity structure.

\subsection{Gravitational Lensing}
Gravitational lensing is proposed to be one of the most promising tools to search for and study DM, as DM itself barely interacts with any electromagnetic (EM) waves ~\cite{Massey_2010}, but the gravitational field generated by DM would still act as a lens and bend the trajectory of light rays traveling nearby~\cite{https://doi.org/10.1002/andp.19163540702}. In this work, we explore the possibility of discovering the aforementioned BEC vortices in a galaxy by gravitational lensing observation.

If the background light source (such as a quasar, supernova or galaxy), the gravitational lens, and the observer are sufficiently aligned, the image(s) of the source would be displaced and magnified (or de-magnified), and their shapes may be distorted because of the deflection of light by the curved spacetime induced by the gravitational lens. There may also be multiple images of the same source arriving at different time ~\cite{https://doi.org/10.1002/andp.19163540702, 1992grle.book.....S}. 

\begin{figure}
\centering
\includegraphics[width=\columnwidth]{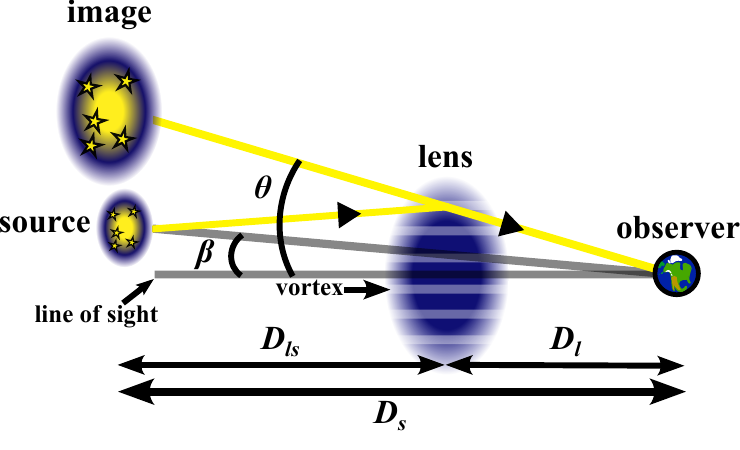}
\caption{Geometry of a gravitational lens system: $D_l, D_s, D_{ls}$ are the (angular diameter) distances from the observer to lens, observer to source, and lens to source, respectively. $\vec{\beta}$ and $\vec{\theta}$ are the angular positions of the source and image, respectively, relative to the line joining the observer and the center of the lens.}
\label{fig:lens_geometry}
\centering
\end{figure}

Fig. \ref{fig:lens_geometry} illustrates the setup of a general gravitational lens system. Here, the source and image angular positions are denoted by $\vec{\beta}$ and $ \vec{\theta}$, respectively. $D_l, D_s, D_{ls}$ denote the angular diameter distances from the observer to the lens, observer to source, and lens to source, respectively. Also, the redshifts of the lens and source are $z_l$ and $z_s$, respectively. The scale of the lens is characterized by the Einstein radius $\theta_E$, which depends on the lens mass and the distances mentioned above. The deflection/lens potential $\psi$ is then defined by
\begin{equation}
    \psi(\vec{\theta})=\frac{1}{\pi}\int{\kappa(\vec{\theta}') \text{ln}|\vec{\theta}-\vec{\theta}'| d^2 \theta'},
\end{equation}
where the convergence $\kappa$ is given by
\begin{equation}
    \kappa(\vec{\theta})=\frac{\Sigma(\vec{\theta})}{\Sigma_{cr}},
\end{equation}
with $\Sigma(\vec{\theta})$ the projected (or column) mass density of the lensing object along the observer's line of sight and $\Sigma_{cr}={c^2 D_s}/{4 \pi G D_l D_{ls}}$ the critical density~\cite{Takahashi_2003, tambalo2023gravitational}.

The lens equation is then
\begin{equation}
    \vec{\beta} = \vec{\theta} - \nabla_{\vec{\theta}} \psi(\vec{\theta}),
    \label{lens eq}
\end{equation}
with which the image position(s) $\vec{\theta}$ can be solved given a known source position and lens model.

The magnification $\mu$ of the image brightness is given by~\cite{Takahashi_2003, Bulashenko_2022}
\begin{equation}
    \mu(\vec{\theta})=\left[\det \left(\frac{\partial \vec{\beta}}{\partial{\vec{\theta}}}\right)\right]^{-1},
    \label{magnification eq}
\end{equation}
where

\begin{equation}
    \frac{\partial \vec{\beta}}{\partial{\vec{\theta}}}=
\begin{pmatrix}
1-\frac{\partial^2 \psi}{\partial \theta_1^2} & -\frac{\partial^2 \psi}{\partial \theta_1 \partial \theta_2}\\
-\frac{\partial^2 \psi}{\partial \theta_2 \partial \theta_1} & 1-\frac{\partial^2 \psi}{\partial \theta_2^2}
\end{pmatrix},
\end{equation}
and $\theta_{1,2}$ are the two components of the vector $\vec \theta$.

Note that the change in brightness of a lensed image is described by the magnitude of $\mu$ and its sign is not important here.

While the above formalism describes the gravitational lensing of a point source, it can also be applied to lensing of any extended source, such as a galaxy, since an extended source is just a collection of point sources. In this work, we will focus on the lensing effect induced by the vortices in the lens galaxy halo, when an extended source such as a background galaxy is lensed. 

Lensing has become a standard, mature and powerful tool in modern astronomy. Its applications span from cosmography to galaxy and galaxy cluster mass profiles study~\cite{Bartelmann_2010, Koopmans_2003}. Moreover, many attempts to probe substructures embedded in a smooth galaxy lens profile were also carried out~\cite{D_az_Rivero_2018}: while the image positions and time delays of the lensed images are dominated by the smooth lens model, their magnifications are sensitive to the (relatively) small mass clumps or substructures embedded in the lens~\cite{Keeton_2003}. Therefore, if substructures are present, the image positions and the magnifications would not agree with each other under a smooth lens model. This so-called 'flux ratio anomaly'~\cite{10.1046/j.1365-8711.1998.01319.x} has been studied extensively ~\cite{10.1046/j.1365-8711.1998.01319.x, Chan_2020, Amruth_2023}, as it could be valuable for probing substructure physics and DM properties~\cite{10.1046/j.1365-8711.1998.01319.x, Keeton_2003, D_az_Rivero_2018}.

Motivated by the flux ratio anomaly, we study the additional lensing effect induced by the vortices, especially since they appear as regularly separated underdensity 'holes' in the lens' projected smooth density profile. 

\section{Simulation}\label{sec:Method}

Since the primary focus of this study is the vortex formation and detection, different angular momentum values were used to observe their effect on the halo profile. Below, we outline the key parameters used in the simulations. 


The initial density profile is chosen to follow the TF profile in Fig.~\ref{fig:TF}, with a core radius of approximately 1 kpc, corresponding to that of a Milky Way-like galactic core. While the choice of the initial profile is not critical — since the simulation evolves towards a state of minimal energy — selecting an initial profile that approximates the final result can significantly reduce the simulation time.


The parameters used in the simulation are $m = 2.92 \times 10^{-22}$ eV, $g=8.26 \times 10^{-22} \text{ eVfm}^3$, which falls within the theoretically predicted values for stable solutions~\cite{Suarez:2016eez}. The mass of the soliton core is chosen to be $6.39 \times 10^{10} M_\odot$. Four values of rotational speed $v_0 = R_0\Omega$ with $R_0 = 2~\text{kpc}$ are used: [0, 40, 66.7, 120] km/s~($\Omega = [0, 1.17, 1.95, 3.51]~\text{deg/Myr}$), assumed to be the same for the baryonic matter.

In Eq.~\ref{eq:GPP}, the baryonic potential $\Phi_{\text{Bary}}$ is included to account for the gravitational effects of baryonic components in the galaxy center, such as the supermassive black hole, stars, and gas. While it has been shown that introducing an external potential can help to stabilize the vortex structures~\cite{Glennon:2023oqa}, the system's sensitivity to the choice of this potential will introduce additional uncertainties. Moreover, the precise mass distribution in the Milky Way center is not well established. Therefore, the baryonic potential is neglected in this study, focusing instead on a pure dark matter halo without baryonic feedback.

The simulation box with periodic boundary condition has side lengths of 10 kpc, ensuring that the boundary effect is negligible.


\section{Results}\label{sec:Result}

\subsection{Simulation Results}

\begin{figure}[t!]
    \centering
    \includegraphics[width=90mm]{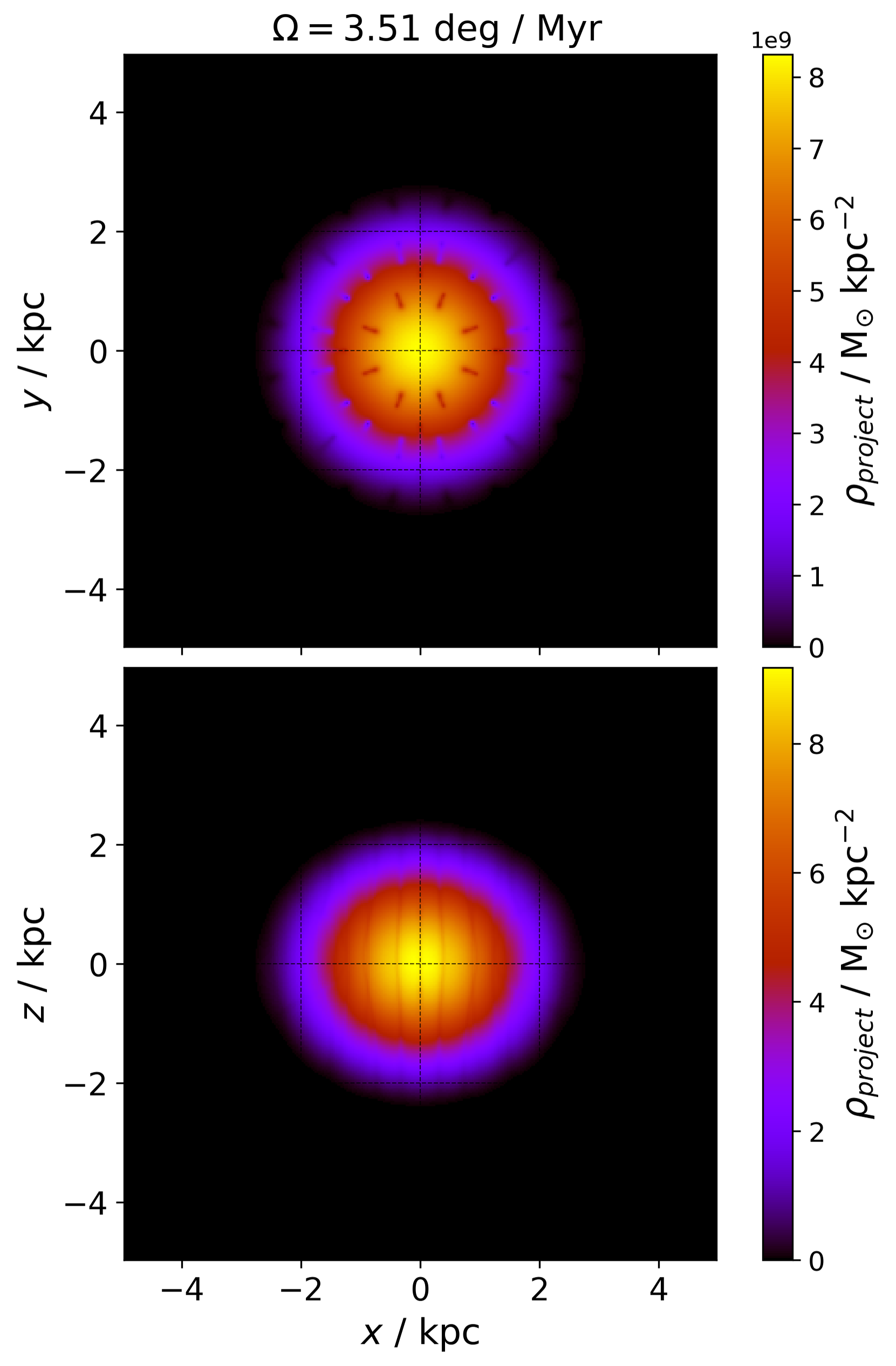} 
    \caption{
        Column density profile for a BEC halo with $\Omega = 3.51$ deg/Myr. The top (bottom) panel shows the top (side) view with the $z$-axis as the rotational axis.
    }
    \label{fig:omega45}
\end{figure}

\begin{figure}[t!]
    \centering
    \includegraphics[width=70mm]{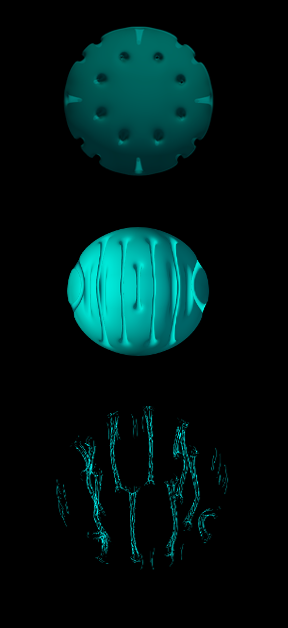} 
    \caption{
        3D illustration of the simulated result in Fig.~\ref{fig:omega45}. Top, side, and vortex-only views are shown from top to bottom.
    }
    \label{fig:3dVortex}
\end{figure}

Fig.~\ref{fig:omega45} shows the column density, $\rho_{project}$, of the lowest energy state of a rotating BEC halo. The angular momentum is computed assuming rigid rotation with an angular speed $\Omega = 3.51$ deg/Myr. Vortices emerge, forming a stack of multiple ring-like structures. The side view (bottom panel) reveals that the DM halo is flattened due to the centrifugal force, while the vortices appear as pillar-like formations penetrating the entire structure. The 3D structures of the same result are shown in Fig.~\ref{fig:3dVortex}. This configuration shows a resemblance to the 3D evolution observed in laboratory-scale experiments~\cite{castin1999bose}.

\begin{figure}[t!]
    \centering
    \includegraphics[width=90mm]{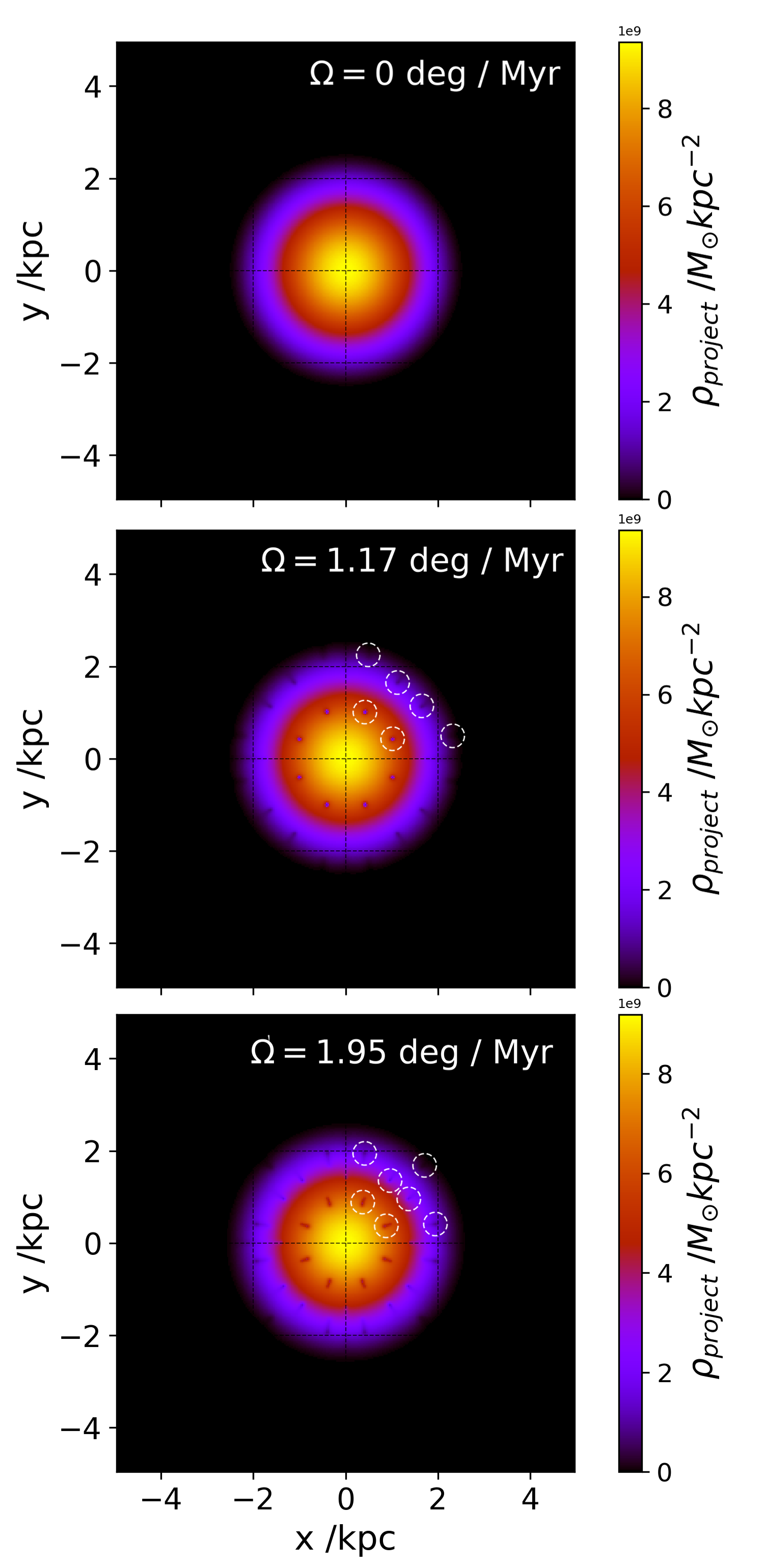}
    \caption{
        Same as Fig.~\ref{fig:omega45}, but for $\Omega = 0 \text{ deg/Myr}$ (top), $1.17 \text{ deg/Myr}$ (middle), and $1.95 \text{ deg/Myr}$ (bottom), respectively. The vortices in the upper right quadrant are outlined with dashed circles to guide the eyes..
    }
    \label{fig:omega0to25}
\end{figure}

Fig.~\ref{fig:omega0to25} presents the simulation results for different values of $\Omega$. As $\Omega$ increases, the number of vortices also increases. This is expected, as more vortices are needed to carry the additional angular momentum. However, beyond a critical value of $\Omega$, the structure becomes unstable and is torn apart by the growing centrifugal force.

\subsection{Gravitational Lensing Result}\label{sec:Lensing_Result}
\begin{figure*}[t!]
    \centering
    \includegraphics[height=1\linewidth]{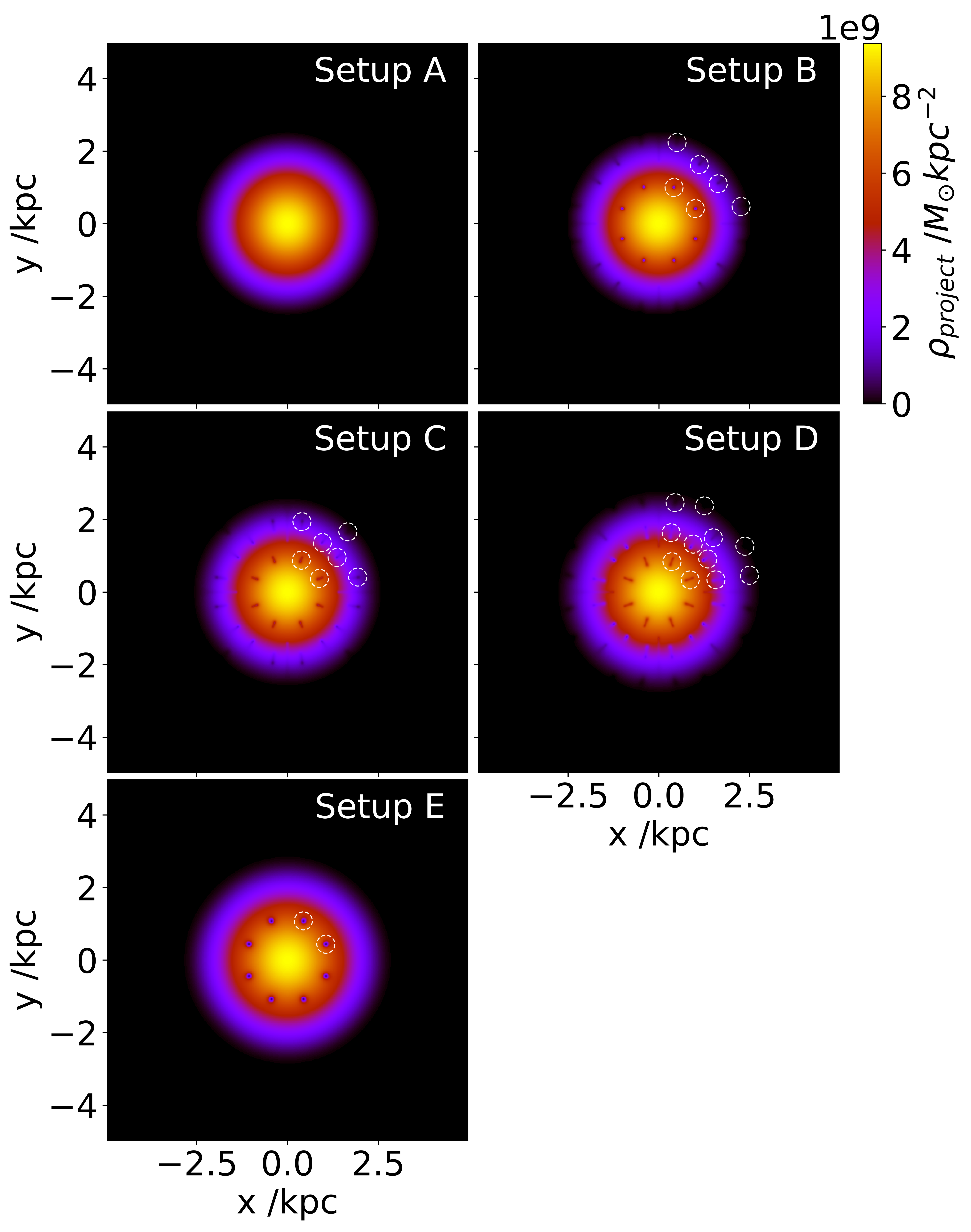}
    \caption{Column density profiles of Setups A to E while the rotational axis is along the line of sight, from left to right, then top to bottom, respectively. As shown in Table \ref{tab:setup_list_table}, Setups A to D share the same total mass and coupling strength $g$, with different angular speeds $\Omega = 0, 1.17, 1.95$ and $3.51$ deg/Myr, respectively, giving rise to increasing under-density holes. Meanwhile, Setup E has $\Omega = 1.56$ deg/Myr and shares the same coupling strength $g$ as Setups A to D, but its total mass is $10\%$ of Setup C's. The vortices in the upper right quadrant are outlined with dashed circles to guide the eyes..
     }
    \label{fig:2d density top view}
 \end{figure*}

To examine the lensing effect induced by the vortices, we set up a gravitational lens system. For the lens, we generate BEC cores with different number densities, patterns and sizes of vortices (by varying the DM parameters). We then join the BEC core obtained from our simulation to an NFW (Navarro-Frenk-White) density profile in the outskirt. They are joined by simple interpolation to ensure a smooth and extended density profile suitable for gravitational lensing calculations. For selected profiles, we study two simple cases: 1. the halo's rotational axis is along the line of sight, so that the vortices will appear as under-density 'holes' in the projected density profile; 2. the rotational axis is perpendicular to the line of sight, such that the vortices will appear as under-density 'strips' in the column density.

\begin{table*}
    \centering
    \begin{tabular}{|c|c|c|c|c|c|}
        \hline
        Setup & A & B & C & D & E\\
        \hline
        Total mass (in $M_\odot$) & $6\times 10^{10}$ & $6\times 10^{10}$ & $6\times 10^{10}$ & $6\times 10^{10}$ & $6\times 10^{9}$\\
        \hline
        Angular speed $\Omega$ (in deg/Myr) & 0 & 1.17 & 1.95 & 3.41 & 1.56 \\
        \hline
    \end{tabular}
    \caption{Parameters for simulating the lens profiles in Setups A to E. The corresponding column density of each profile is shown in Fig.~\ref{fig:2d density top view}. All setups have the same coupling strength $g=8.26 \times 10^{-22} \text{ eVfm}^3$.}
    \label{tab:setup_list_table}
\end{table*}

The lens profiles used in the simulation are summarized in Table \ref{tab:setup_list_table}, which are denoted as Setups A to E. The column densities of these lens profiles while the rotational axis is along the line of sight are also presented in Fig.~\ref{fig:2d density top view}. Setups A to D share the same total mass of $6\times 10^{10} M_\odot$ and coupling strength $g=8.26 \times 10^{-22} \text{ eVfm}^3$, and they have increasing angular speeds of $0, 1.17, 1.95$ and $3.51$ deg/Myr, respectively. Since Setup A is non-rotating, it does not have any vortices, and it serves as a reference for examining the effect induced by the vortices. For Setups B to D, while the positions and numbers of the innermost under-density holes (in the column density) caused by the vortices remain roughly the same, as the rotational speed increases, the sizes of the holes also increase, and the holes are stretched radially. Setup E has a rotational speed of $1.56$ deg/Myr and shares the same coupling strength as Setups A to D, but its total mass is only 10\% of the latter. As a result, the under-density holes created by the vortices are deeper. 

In order to study the lensing effect of the vortices, the source (background) galaxy and the lens are placed at redshift $z_s=0.7$ and $z_l=0.5$, respectively, and then we set up the source position such that the lensed images are located close to the vortices on the lens plane. We assume the standard Planck18 cosmology in the calculation~\cite{2020}. 

For the source galaxy, we set up an intermediate size, NGC4725-like galaxy~\cite{10.1093/mnras/stt1320} described by a Sérsic light profile with a half-light radius of $R_e=14.9$ kpc and Sérsic index $n=4.14$. For simulating the lensed images, we assume the point spread function (PSF) of the telescope to be a Gaussian with full width at half maximum (FWHM) $\sigma = 1, 10$ and $70$ milliarcseconds, respectively, with the first two roughly representing that for VLBI (Very-long-baseline interferometry) at radio band~\cite{2020arXiv200702347V} and the last one the Hubble space telescope at visible light range~\cite{Windhorst_2011}. As we will show later, the width of the PSF has an important effect on observing the fine details of the lensed images induced by the vortex. We note that, realistically, the light profile of a galaxy is different for radio and visible light observations, as the visible light component is dominated by stars, gas clouds, interstellar dust (and possibly active galactic nuclei) in the bulge and disc components~\cite{10.1093/mnras/234.1.131}, while the radio component is contributed by processes such as synchrotron radiation (for example, if there is an active galactic nucleus or star forming region)~\cite{La_Mura_2017, 10.1093/mnras/stac1808}, emission from H II regions~\cite{1997A&A...323..323D} and neutral hydrogen gas (21-cm line)~\cite{Dutta_2022}. In general, depending on the galaxy's type and structure, the radio emission region may still have a similar or even larger size compared to that of visible light, despite different shapes and profiles. For instance, extended radio lobes from active galactic nuclei, such as that in the radio galaxy Centaurus A~\cite{2004igc_book_J, Hardcastle_2003}, have a larger emission region. Therefore, in practice, for the three cases of different $\sigma$'s, the light profiles' shapes and sizes would not be the same. However, we neglect this detail and assume the same Sérsic light profile across the three cases, since we focus on the effects induced by lensing. In addition, we also neglect any light from the lens galaxy and noises, representing an ideal optimal situation.

\begin{figure*}[t!]
    \centering
    \includegraphics[height=.3\linewidth]{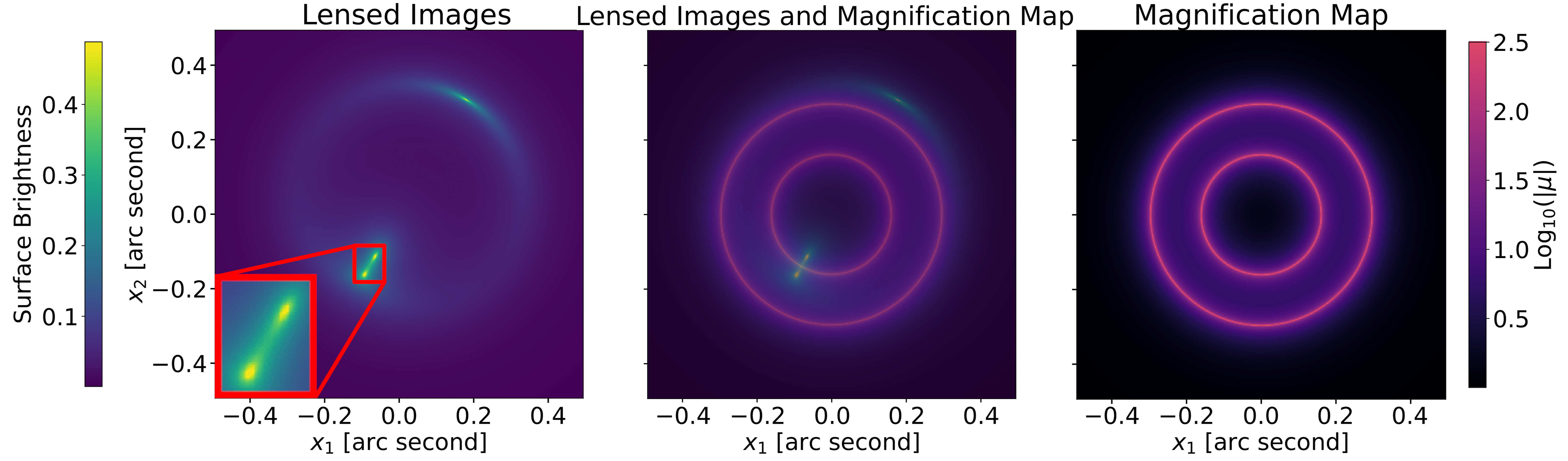}
    \caption{Lensing result of a profile without vortices (Setup A). Lens and source redshifts are $z_l=0.5$ and $z_s=0.7$, respectively. Left: lensed images (with PSF's FWHM set to $1$ milliarcseconds), with an enlarged snapshot of the region bounded by red rectangle. Middle: overlay of lensed images and magnification map. Right: magnification map of the lens profile. For the surface brightness shown in the image, we note that the unit is arbitrary, as we only need to focus on the relatively brightness on different parts of the images.
     }
    \label{fig:setup A result}
 \end{figure*}
 
\begin{figure*}[t!]
    \par\medskip
    {
    \centering
    \includegraphics[height=.75\linewidth]{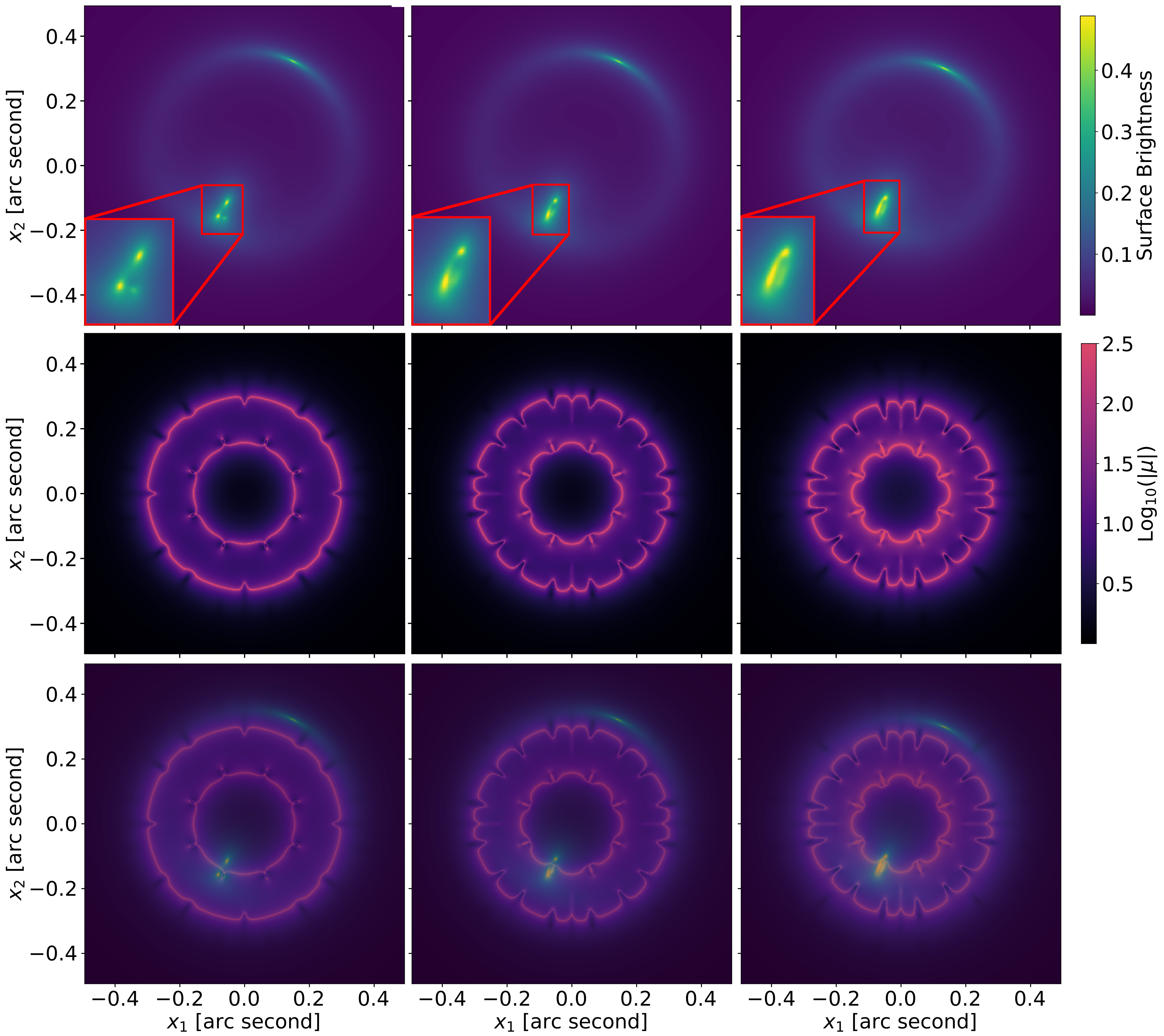}
    }\par

    \caption{Lensing results of Setups B, C and D shown from left to right, which have the same halo mass but different angualar speeds of 1.17, 1.95, and 3.41 deg/Myr, respectively. Lens and source redshifts are $z_l=0.5$ and $z_s=0.7$, respectively. The top panels show the lensed images of each setup (with $\sigma$ set to $1$ milliarcseconds). The middle panels show the magnification map of each lens profile, and the bottom panels show the overlay of lensed images and magnification map. For the surface brightness shown in the image, we note that the unit is arbitrary, as we only need to focus on the relatively brightness on different parts of the images.
     }
    \label{fig:setup B C D result}
 \end{figure*}

In Fig.~\ref{fig:setup A result}, we show the lensing result when there are no vortices, hence no under-density holes in the lens profile. From the simulated image, there are three lensed images of the source/background galaxy. In addition, since the source galaxy is close to the center of the lens and is an extended source, the lensed images are extended. As a result, the two outermost and opposite images are connected by a faint ring/arc. We note that the radius of the ring is similar to the Einstein radius of the lens. Furthermore, from the magnification map and the overlay, we can see that two of the images are located near a critical curve, i.e, the curve with (theoretically) infinite magnification and where images merge on the lens/image plane. Most importantly, the critical curve here forms a very smooth circle in this profile, which is expected, as the lens profile is smooth. 

Fig.~\ref{fig:setup B C D result} shows the simulated lensing results of Setups B to D when the rotational axis is aligned with the line of sight, in which the lensed images with $\sigma = 1$ milliarcseconds are shown in the upper panel, the magnification map shown in the middle panel, and the overlay of the lensed images and magnification map shown in the lower panel. Firstly, for all three images, we can clearly see a bump in the brightness of the lower left image, despite the bumps having different shapes. From the magnification map and the overlay, we can see that the brightness bumps are caused by the distorted critical curve, especially the small loops of high magnification curve which are induced by the vortices. The level of distortion also increases with the vortex size, as we go from Setups B to D. In addition, we note that only one vortex is inducing observable effect, as other vortices are too far away from the images. 

Secondly, around the small loops, there are also regions where the magnification is lower or higher than the surrounding. For instance, for Setup B, the regions that are inside the loop and tangentially next to the loop are demagnified, while the regions that are radially outward next to the loop are magnified. However, these effects may not be easy to detect, since the source galaxy has a limited size and does not have a uniform brightness distribution. It is difficult to visually construct the signatures of the magnification map around the vortex in details. The feasibility of doing this with lens modeling would be left for future studies. Moreover, we note that an over-density region, such as a perturbation caused by subhalos, will also induce anomalies in the image brightness~\cite{oh2024improvingfluxratioanomaly}. It will be important for a future study to identify ways to distinguish if the magnification anomalies are caused by underdensity 'holes' (due to vortices) or overdensity subhalos. 


\begin{figure*}[t!]
    \par\medskip
    {
    \centering
    \includegraphics[width=1\linewidth]{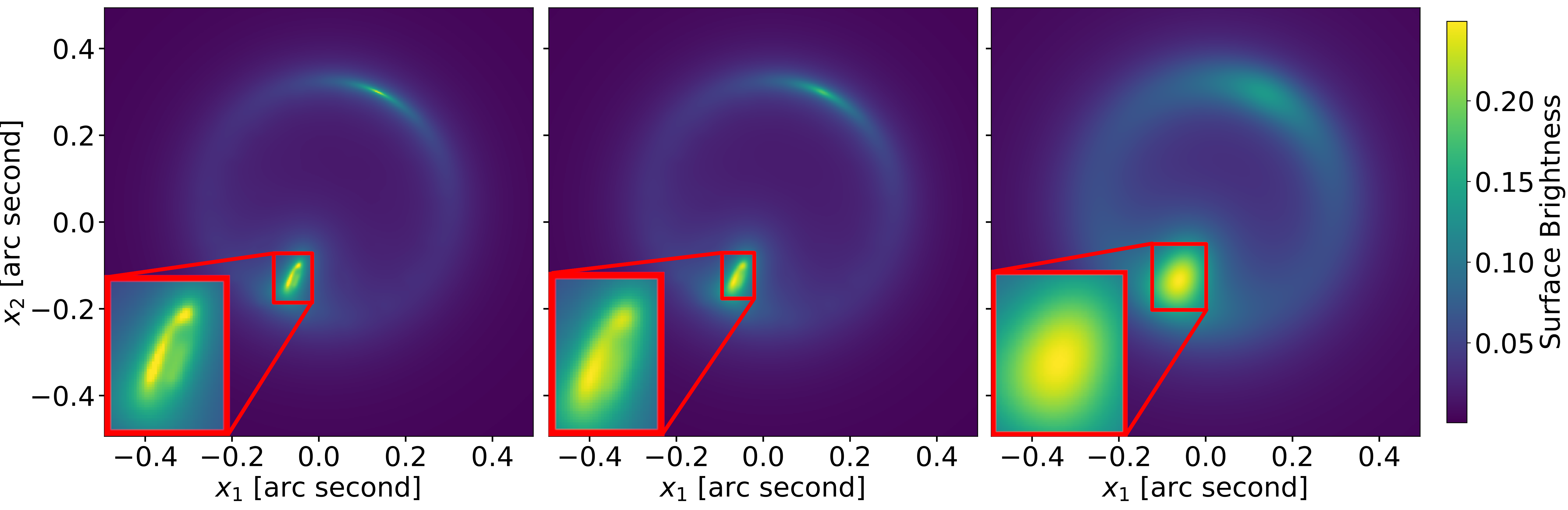}
    }
    
    \caption{Lensed images of Setup D (lens and source redshifts are $z_l=0.5$ and $z_s=0.7$, respectively) with $\sigma = 1, 10$ and $70$ milliarcseconds, shown from left to right, respectively. For the surface brightness shown in the image, the unit is arbitrary, as we only need to focus on the relatively brightness on different parts of the images.
     }
    \label{fig:setup D fwhm comparison}
 \end{figure*}

In terms of observation, for each of the lensed images, we also note that the telescope PSF's FWHM would affect the ability to distinguish the anomalies or signatures induced by the vortices. To illustrate, we compare the lensed images of Setup D at $\sigma$ of $1, 10$ and $70$ milliarcseconds in Fig.~\ref{fig:setup D fwhm comparison}. We can see that, while the brightness anomaly is very clear for $\sigma = 1$ milliarcsecond and still observable for $\sigma = 10$ milliarcsecond, it becomes indistinguishable for $\sigma = 70$ milliarcsecond, as the PSF is wide enough to smooth out the bump in the brightness. Therefore, telescopes with a small enough PSF width, such as VLBI, would be required to observe such lensing signatures caused by vortices. 

\begin{figure*}[t!]
    \par\medskip
    {
    \centering
    \includegraphics[height=.75\linewidth]{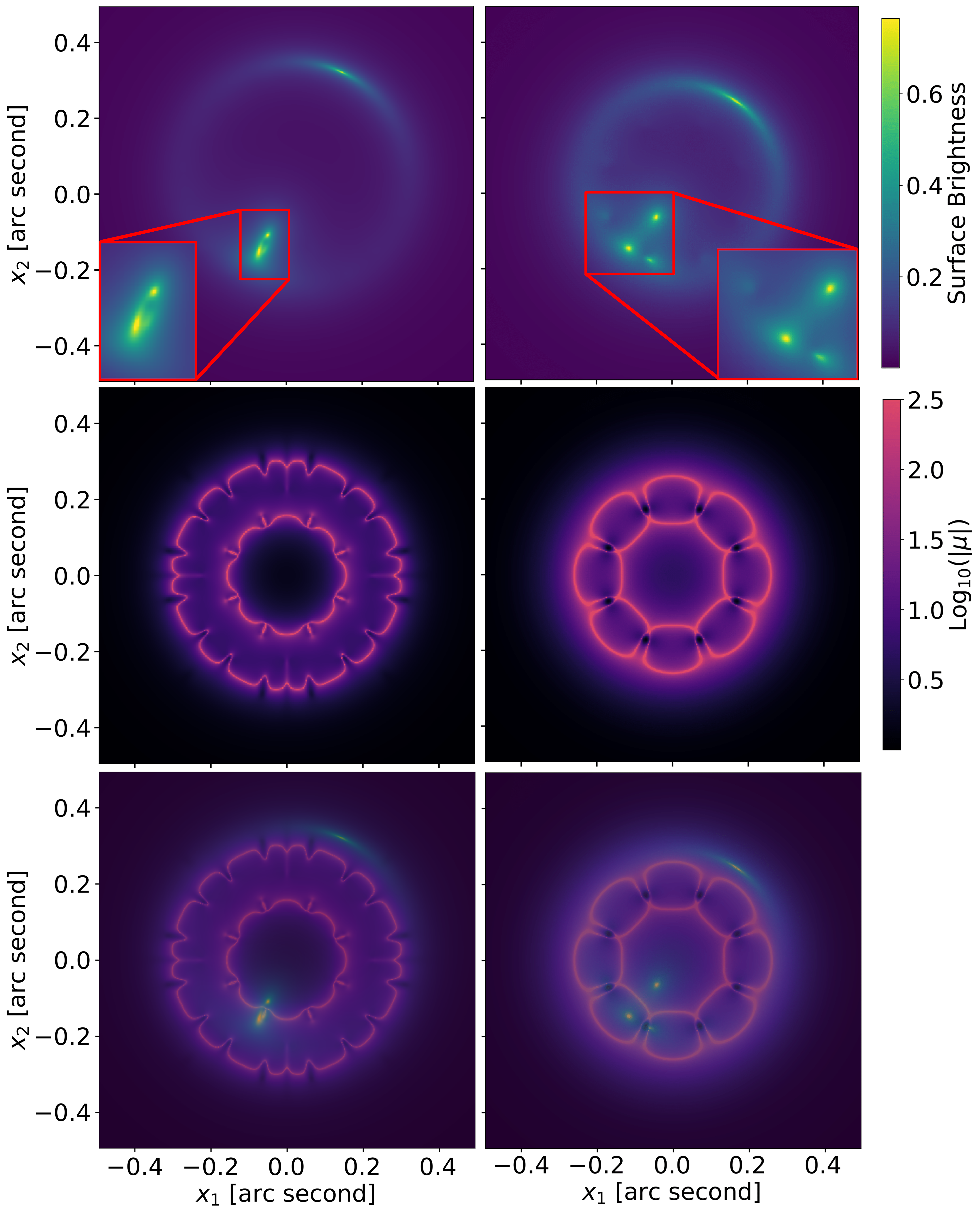}
    }\par

    \caption{Same as Fig.~\ref{fig:setup B C D result}, but for Setups C (left panels) and E (right panels), which have similar rotational speed and same coupling strength but different halo masses and vortex sizes.
     }
    \label{fig:setup C and E result}
 \end{figure*}

In Fig.~\ref{fig:setup C and E result}, we compare the lensing simulations for Setups C and E, which have different total masses but similar rotational speed and the same coupling strength. Firstly, from the projected lens profile in Fig.~\ref{fig:2d density top view}, we can see that Setup E has larger vortices. Its effect is reflected in the magnification map, showing a much greater distortion level of the critical curve than that for Setup C. Hence, the lower corresponding lensed image also has a larger brightness anomaly. Secondly, as the vortices are larger, their effect also become significant on the long arcs that connect the two outermost and opposite images: from the lensed image in Setup E, we can see multiple regularly separated bumps in the ring's brightness, while in Setup C, such features cannot be observed. We note that, in this specific example, the smaller Einstein radius of the lens in Setup E (due to its smaller mass) may also help, as it allows the arc of the image to overlap with more vortices. Hence, a lens profile with larger vortices may not only lead to larger brightness anomalies, but also produce anomalies that are regularly separated, which provide a strong evidence for BEC ULDM. 

\begin{figure*}[t!]
    \par\medskip
    {
    \centering
    \includegraphics[height=.75\linewidth]{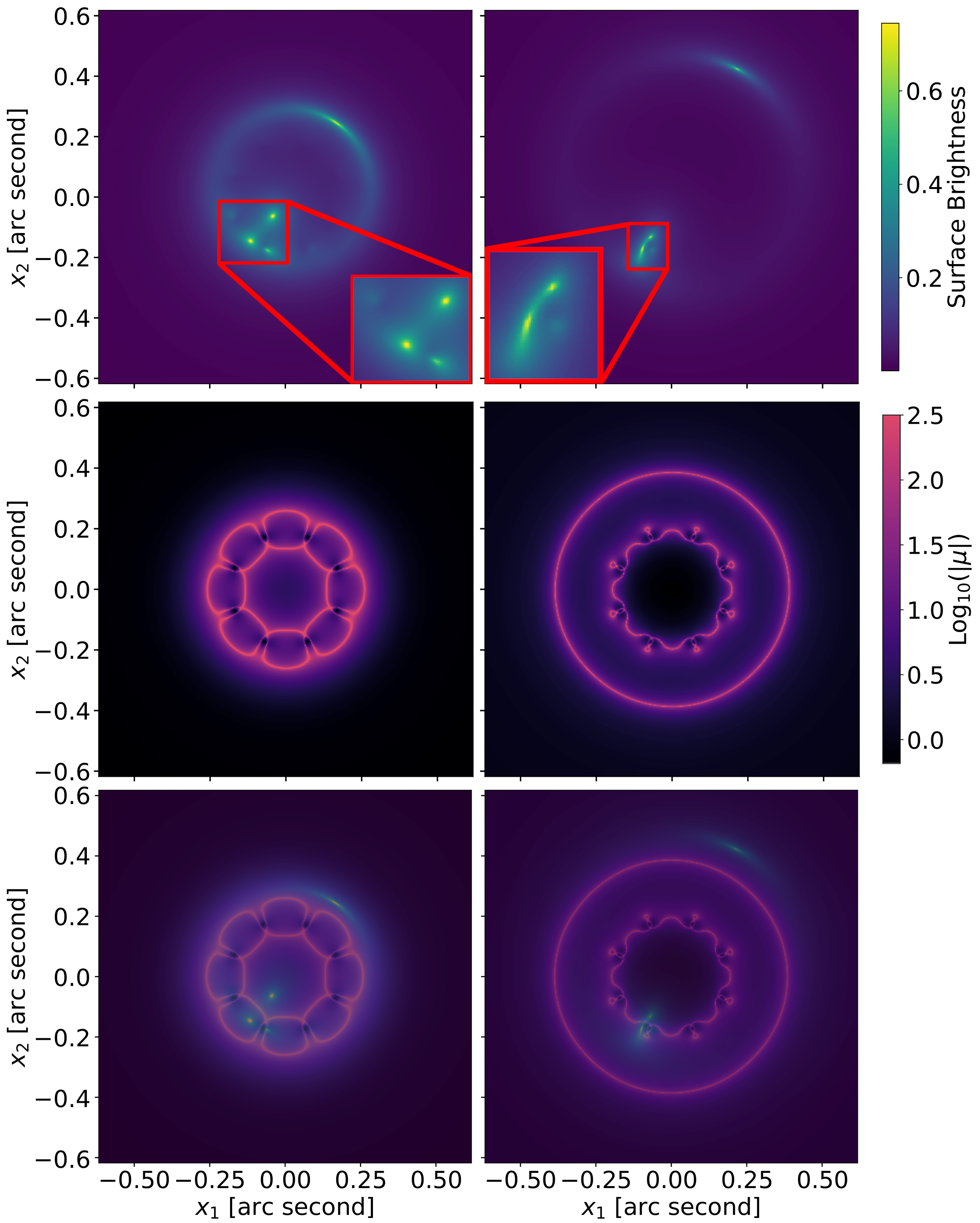}
    }\par

    \caption{Same as Fig.~\ref{fig:setup C and E result}, but for Setup E at source redshifts of $z_s=0.7$ (left panels) and $z_s=0.9$ (right panels), respectively
     }
    \label{fig:setup E different zs comparison}
 \end{figure*}

The effect of the vortices on lensed images also depends on the distances in the observer-lens-source system. In Fig.~\ref{fig:setup E different zs comparison}, we compare the lensing result of Setup E with the source located at $z_s=0.7$ and $0.9$, respectively, while the lens red shift remains at $z_l=0.5$. We observe that the distortion of the critical curve and the lensing signature of the vortex on the lensed images are reduced for the $z_s=0.9$ case. In addition, similar to Setup C shown in Fig.~\ref{fig:setup C and E result}, the effects of other vortices also become insignificant. In short, the vortices' effects on the resulting lensed images depends on not only the lens profile, but also other external parameters, such as the position and redshift of the source galaxy.

\begin{figure*}[t!]
    \centering
    \includegraphics[height=.28\linewidth]{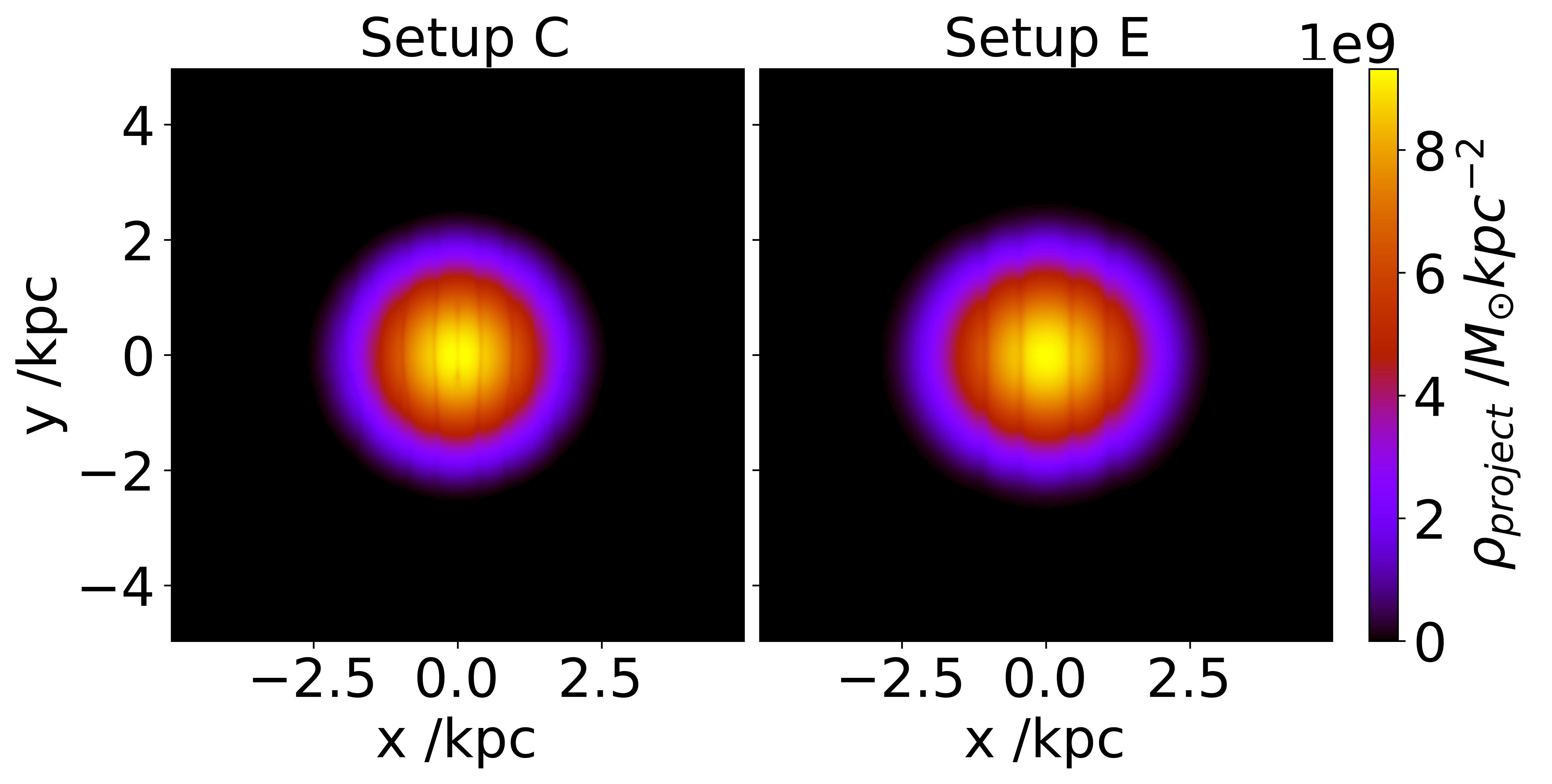}
    \caption{Projected lens profile of Setups C and E with the rotational axis being perpendicular to the line of sight, on left and right panels, respectively.
     }
    \label{fig:2d density side view}
 \end{figure*}

\begin{figure*}[t!]
    \par\medskip
    {
    \centering
    \includegraphics[height=.75\linewidth]{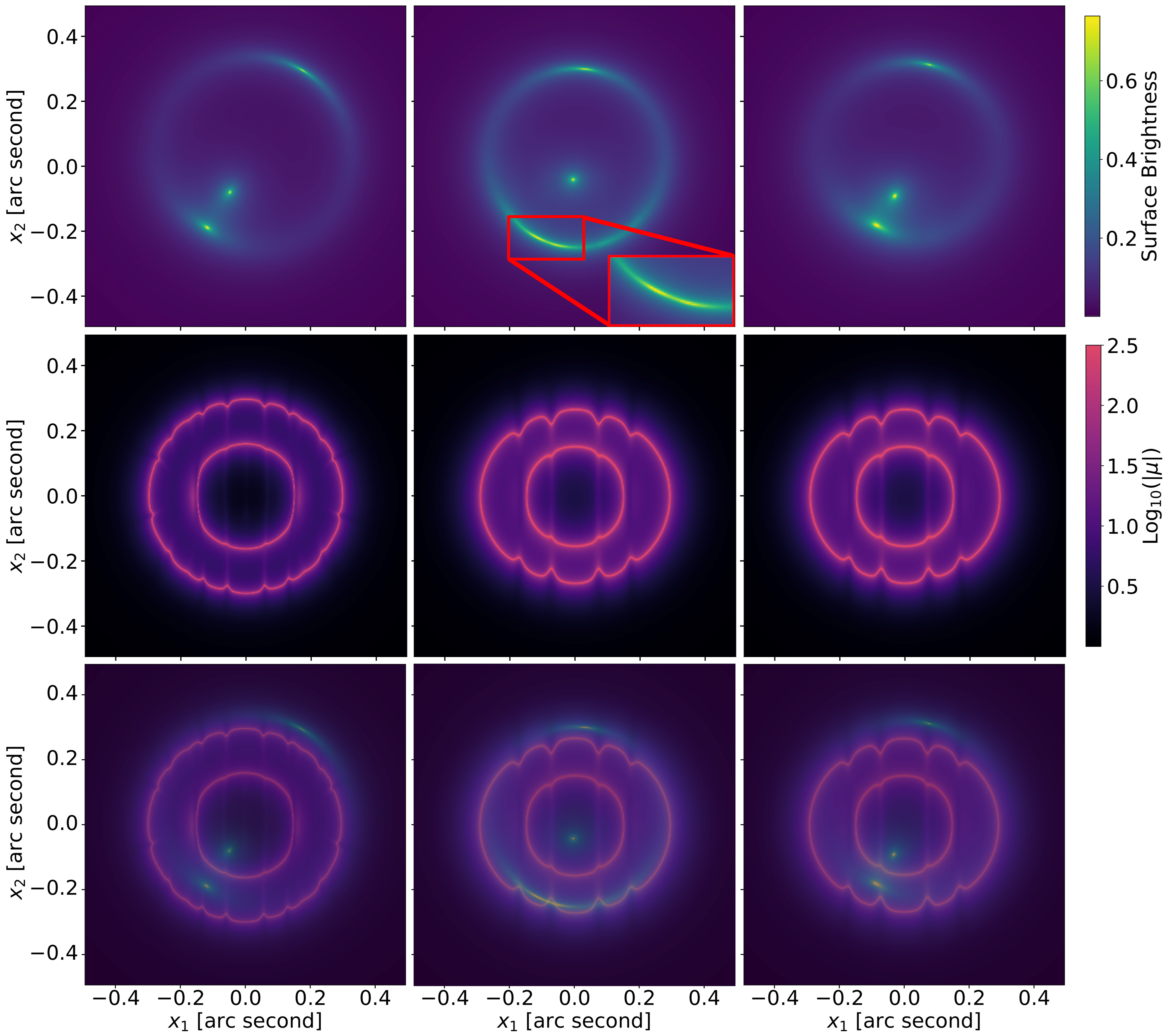}
    }\par

    \caption{Same as Fig.~\ref{fig:setup C and E result}, but with the rotational axis of the lensing halo perpendicular to the line of sight. The middle and right columns both correspond to results of Setup E but with their source positions placed differently (the setup in the middle panel has a source galaxy aligned closer to the center of the lens). The left column corresponds to result of Setup C. 
     }
    \label{fig:side view result}
 \end{figure*}

In an actual observational scenario, it would be very probable that the rotational axis is not along the line of sight of the observer and the vortices will appear as under-density strips rather than holes. Hence, the lensing effect of the vortices presented above represents the optimal scenario. When the rotational axis is perpendicular to the line of sight, the lensing signature of the vortex can be very insignificant. In Fig.~\ref{fig:2d density side view} and Fig.~\ref{fig:side view result}, we present the column densities of Setups C and E viewed in such an orientation and their corresponding lensing simulation results, respectively. 

From Fig.~\ref{fig:2d density side view}, we can see that the under-density strips created by the vortices are not as significant in density variations as the under-density holes (Fig.~\ref{fig:2d density top view}). This is because the normal density parts of the lens compensate for the under-density of the vortex during the integration of density along the line of sight. For Setup E, we present two different cases (in the middle and right panels of Fig.~\ref{fig:side view result}) where the source galaxies are placed in two different positions (with the case in the middle panel having a source galaxy aligned closer to the center of the lens). In particular, for the case presented in the middle panel of Fig.~\ref{fig:side view result}, the lower lensed image overlaps with the distorted critical curve. This creates a boost in the brightness, then a reduction, followed by a boost again on the lower image tangentially along the critical curve. However, for Setup E in the right panel and Setup C in the left panel where the lensed images only overlap with the strip of perturbed magnification, the lensing effects of the vortices are not significant. The non-uniform brightness of the source galaxy light profile also increases the difficulty to spot such effects.

To summarize, the vortices in the lens galaxy would induce anomalies in the flux of the lensed images if they are located near the images. The strength of the anomaly will be affected by the sizes of the vortices, as well as the alignment between the halo's rotational axis and the observer's line of sight. A higher number density of vortices would also increase the chance that the lensed images land near a vortex. Moreover, as demonstrated in some scenarios, the lensed image of the background galaxy may also overlap with multiple vortices, which would be a strong indication of the existence of such structures. In short, we have shown that lensing offers a possible observational method to discover vortex structures in ULDM halo. More detailed studies on the feasibility and limitation of lensing observation of BEC vortices will be carried out in the future.

\section{Conclusions and Outlook}\label{sec:Conclusions}
We have demonstrated, using a numerical solution of the GPP equation, that a lattice of vortices — underdensity columns that carry angular momentum — naturally forms in a ULDM halo under conditions similar to those of the Milky Way, such as a rotational speed of around 40-120 km/s at 2 kpc from the halo center. The number, spacing, and size of vortices depend on the mass and coupling strength of ULDM particles, as well as the halo mass and rotational speed. Therefore, the presence of these vortices would be a distinctive signature of BEC-ULDM. It will be important to find ways to distinguish these underdensity vortices from overdensity substructures. We propose to use gravitational lensing as a tool for observational searches of BEC-ULDM vortices. We show that these vortices would induce flux anomalies in the lensed images. In optimal scenarios — characterized by a large vortex size and alignment of the halo's rotational axis with the line of sight — regularly separated brightness anomalies can be produced, providing strong evidence for the existence of the vortex lattice, and thus also for BEC-ULDM.

\section*{Acknowledgments}

This research is supported by grants from the Research Grants Council of the Hong Kong Special Administrative Region, China, under Project No. AoE/P-404/18 and 14300223.

\clearpage
\bibliography{bib.bib}

\newpage
\appendix{Appendix A: Numerical algorithms and data structure}

\subsection{Stationary State Finder}

We are interested in the stationary state solution of Eq.s 4 and 5 describing a self-gravitating BEC, which can be regarded as a constrained minimization problem
\begin{equation}
\phi\in\underset{\|\phi\|=1}{\operatorname*{\arg\min}}\mathcal{E}_{tot}(\phi),
\end{equation}
where $\phi=\Psi/\sqrt{n}$ is the dimensionless single particle wave function and $\mathcal{E}_{tot}$ is the corresponding dimensionless total energy. As explained by Atonie~\cite{antoine_efficient_2017}, the normalization condition ensures that all possible wave functions form a spherical manifold $\mathcal{S}=\{\phi\in L^2(\mathbb{R}^d),\|\phi\|=1\}$. The corresponding tangent space of the spherical manifold is defined by $T_\phi\mathcal{S}=\{h\in L^2(\mathbb{R}^d),\mathrm{Re}\left<\phi,h\right>=0\}$. 

The first-order necessary condition for the local minimum states that the projection of the gradient at the local minimum on the tangent space is zero.

\begin{equation}
    \nabla \mathcal{E}-Re<\phi,\nabla \mathcal{E}>\phi=0.
\end{equation}

From the dimensionless total energy
\begin{equation}
    \begin{aligned}\mathcal{E}_{\mathrm{tot}}(\phi)&\begin{aligned}=\int_{\mathbb{R}^3}[\frac{1}{2}|\nabla\phi|^2+(\Phi+\Phi_{Bary})|\phi|^2+(gN/2)|\phi|^4\end{aligned}\\&\begin{aligned}-\mathrm{Re}(\phi^*{\Omega}{L_z}\phi)]d\mathbf{x}\end{aligned}\\&:=\mathcal{E}_{\mathrm{kin}}(\phi)+\mathcal{E}_{\mathrm{pot}}(\phi)+\mathcal{E}_{\mathrm{int}}(\phi)+\mathcal{E}_{\mathrm{rot}}(\phi),\end{aligned}
\end{equation}
we have the corresponding gradient flow
\begin{equation}
    \nabla\mathcal{E} (\phi)=2\left[-\frac{1}{2}\Delta+V(\mathbf{x})+gN(|\phi|^2)-{\Omega}{L_z}\right]\phi=2H_\phi\phi.
\end{equation}

Therefore, the necessary condition for a local minimum is

\begin{equation}
    \begin{aligned}
       M_\phi(\nabla \mathcal{E})&=2H_\phi\phi -Re<\phi,2H_\phi\phi >\phi\\&=2H_\phi\phi -2\lambda\phi=0,
    \end{aligned}
\end{equation}
where $M_\phi$ is the projection operator for $\mathcal{S}$ and $\lambda=<H_\phi\phi,\phi>$ is the Lagrange multiplier associated with the spherical constraint. 

The traditional method to locate the local minimum is called the projected gradient flow method (imaginary time method). During the imaginary-time evolution, where $\tau=it$, 
\begin{equation}
    \partial_\tau\phi=-(H_\phi\phi-\lambda(\phi)\phi)=-\frac{1}{2}M_\phi\nabla E(\phi).
\end{equation}
The eigenmodes damp, thus decreasing the energy, until the wavefunction no longer evolves.

However, in our code, we used the preconditioned conjugate gradient method instead. The logic is similar, minimizing the energy following the gradient flow, but it is computationally cheaper and more efficient.

\subsubsection{Pseudo-spectral Discretization}
We follow Antonie's method~\cite{antoine_efficient_2017} to implement the code using a standard pseudo-spectral discretization based on Fast Fourier Transforms (FFTs) in a cubic domain of length $L$ with periodic boundary condition and evenly spaced grid size $h=2L/(M-1)$. The corresponding Fourier frequencies $(\xi_p,\mu_q, k_r)$ are $\xi_p=p\pi/L$, $-M/2\le p\le M/2-1$, $\mu_q=q\pi/L$, $-M/2\le q\le M/2-1$ and $k_r=r\pi/L$, $-M/2\le r\le M/2-1$. 

With this descretization, we have the pseudospectral wavefunction $\widetilde{\phi}$ 

\begin{equation}
\begin{aligned}
    \widetilde{\phi}(t,x,y ,z)&=\frac{1}{M}\sum_{p=-M/2}^{M/2-1}\widehat{\widetilde{\phi}_p}(t,y,z)e^{i\xi_p(x+L)}\\&=\frac{1}{M}\sum_{q=-M/2}^{M/2-1}\widehat{\widetilde{\phi}_q}(t,x,z)e^{i\mu_q(y+L)}\\&=\frac{1}{M}\sum_{r=-M/2}^{M/2-1}\widehat{\widetilde{\phi}_r}(t,x,y)e^{i\mu_r(z+L)},
    \end{aligned}
\end{equation}
and corresponding Fourier coefficients

\begin{equation}
    \begin{aligned}
        \widehat{\widetilde{\phi}_p}(t,y,z)=&\sum_{k_1=0}^{M-1}\widetilde{\phi}(t,x_{k_1},y,z)e^{-i\xi_p(x_{k_1}+L)},\\\widehat{\widetilde{\phi}_q}(t,x,z)=&\sum_{k_2=0}^{M-1}\widetilde{\phi}(t,x,y_{k_2},z)e^{-i\mu_q(y_{k_2}+L)},\\\widehat{\widetilde{\phi}_r}(t,x,y)=&\sum_{k_3=0}^{M-1}\widetilde{\phi}(t,x,y,z_{k_3})e^{-i\kappa_r(z_{k_3}+L)}.
    \end{aligned}
\end{equation}

The single particle Hamiltonian operator $H_\phi$ is then discretized 
\begin{equation}
    H_\phi=-\frac{1}{2}[[\Delta]]+[[V]]+gN[[|\phi|^2]]-\Omega[[L_z]],
\end{equation}
where $[[\Delta]]:=[[\partial_x^2]]+[[\partial_y^2]]+[[\partial_z^2]]$, $[[V]]_{k_1,k_2,k_3}=\Phi(x_{k_1},y_{k_2},z_{k_3})+\Phi_{Bary}(x_{k_1},y_{k_2},z_{k_3})$, $[[\mathbb{L}_z]]:=-i(x[[\partial_y]]-y[[\partial_x]])$.

By FFTs, the differential operator can be diagonalized. Here we illustrate with 


\begin{equation}
  \begin{aligned}
  ([[\partial_x^2]]\widetilde{\phi})_{k_1,k_2,k_3}:&=-\frac{1}{M}\sum_{p=-M/2}^{M/2-1}\xi_p^2
  \widehat{(\widetilde{\phi})}_p(t,y_{k_2},z_{k_3})\\[6pt]
  &\quad\times e^{i\xi_p(x_{k_1}+L)},\\
  (y[[\partial_x]]\widetilde{\phi})_{k_1,k_2,k_3}:&=\frac{1}{M}\sum_{p=-M/2}^{M/2-1}i y_{k_2}\xi_p
  \widehat{(\widetilde{\phi})}_p(t,y_{k_2},z_{k_3})\\[6pt]
  &\quad\times e^{i\xi_p(x_{k_1}+L)}.
  \end{aligned}
\end{equation}

Thus, $H_\phi$ is turned into a sparse diagonal matrix. 

\subsubsection{Numerical Implementation}

This system is solved iteratively using a quasi-static framework. The process alternates between solving the Poisson equation for the gravitational potential and evolving the GPE for the condensate wavefunction until convergence is achieved. We develop the code from BEC2HPC~\cite{gaidamour_bec2hpc_2021}.

The Poisson part is solved trivially with the pseudo-spectral method after FFTs. The updated gravitational potential $\Phi$ is then used as an input for the Gross-Pitaevskii equation. The GPE part is evolved quasi-statically to compute the ground-state wavefunction. The evolution locates the ground state solution numerically using the preconditioned conjugate gradient method, which is well-suited for large and sparse systems.

To prevent numerical drift caused by rotation or boundary effects, the center of mass (COM) of the condensate is recalculated and re-centered at each iteration. The COM is determined as:

\begin{equation}
    \mathbf{R}_{\mathrm{COM}}=\frac{\int\mathbf{r}|\phi|^2d^3r}{\int|\phi|^2d^3r},
\end{equation}

The wavefunction $\phi$ is then shifted to ensure that the COM remains at the origin of the computational domain. This step ensures that the system remains physically consistent and prevents numerical artifacts, particularly in the presence of rotation.

As we mentioned above, periodic boundary conditions are applied to the computational domain, which is large enough to simulate an effectively isolated system. However, under some parameters where the system is highly sensitive to numerical errors (such as low spatial resolution or high rotational speed), an artificial spherical infinite wall could be imposed as a hard-wall boundary condition to prevent density leakage and stabilize the simulation. Tests were performed by varying the wall radius to confirm that the density profile and vortex positions are insensitive to the wall size. We will illustrate this later.

\subsubsection{Parallelization}
The parallel design of the code follows BEC2HPC, which uses the decomposition of the 1D slab in the X direction~\cite{gaidamour_bec2hpc_2021} avoids the interprocessor communication for $[[\partial_y]]$ and $[[\partial_z]]$, while calculation is performed in each slab simultaneously. $[[\partial_x]]$ unavoidably involves a permutation of the axis, which is parallelized internally by the FFTW library~\cite{1386650} with an all-to-all communication.

Another crucial part of parallelization is re-centering, where the mesh is shifted cyclically along the Y and Z directions, while communication is performed between different slabs along the X direction by an Open-MPI send-receive structure.

 We keep the $\boldsymbol{Map}$ objects in BEC2HPC, which not only contain the information of the block distribution along each dimension, but also allow for parallelly creating, accessing, and performing pointwise operations of multidimensional arrays.

\subsection{Convergence Test}
To verify the numerical robustness of our simulations, we performed convergence tests for both non-rotating and rotating self-gravitating BECs.

\subsubsection{Non-Rotating Halo}
\begin{figure}[t!]
    \centering
    \includegraphics[width=90mm]{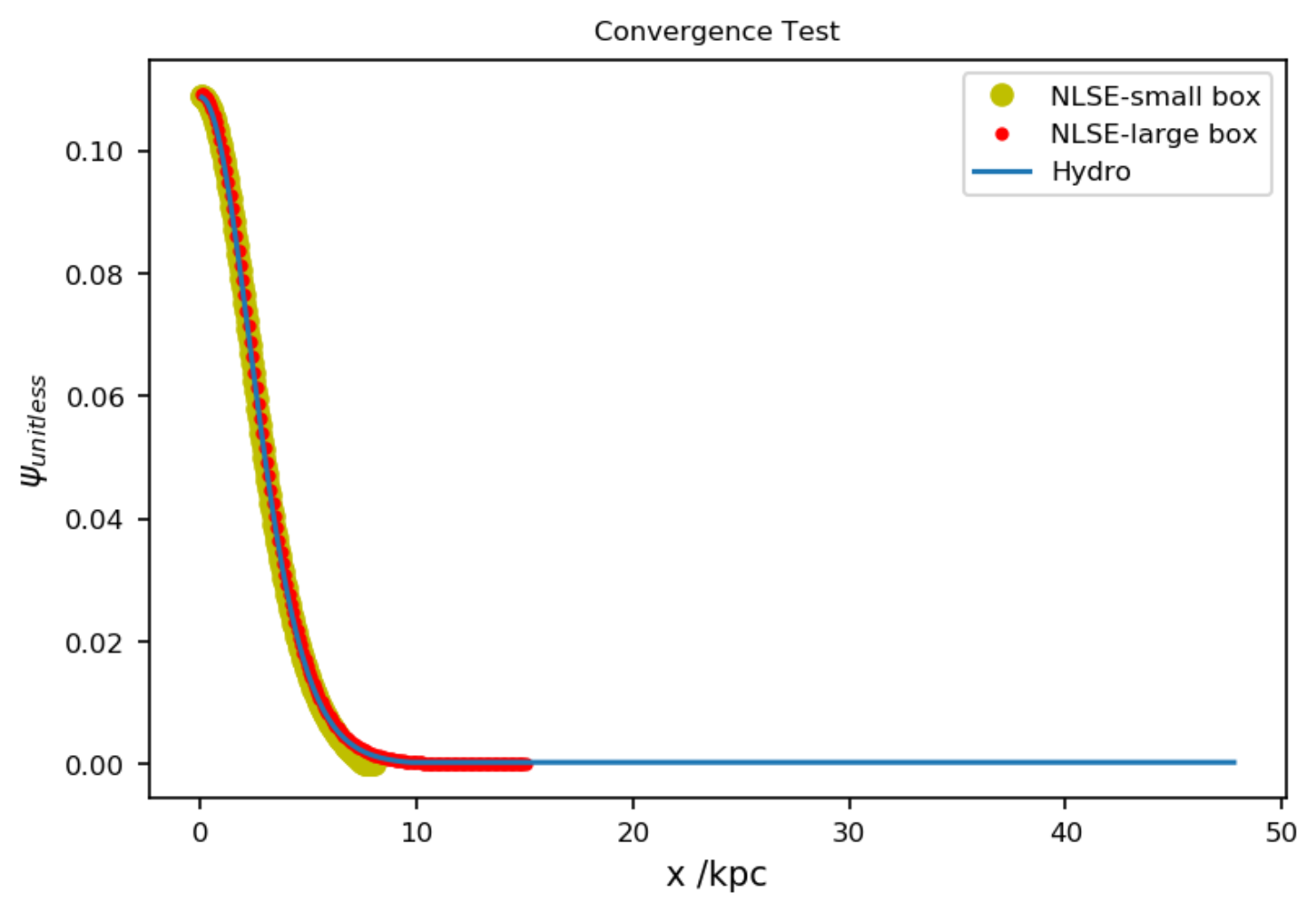}
    \caption{
        Convergence test for the non-rotating halo with respect to grid resolution. The numerical results also agree well with the hydrodynamics result (blue line).
    }
    \label{fig:convnonrotating}
\end{figure}

For the non-rotating case, we tested the convergence of the density profile with respect to grid resolutions. We did simulations with the same physical parameters and number of grid points, but varying the size of the simulation boxes. The results were compared to the corresponding profile obtained from the fluid approach, which serves as an independent benchmark. As shown in Fig.~\ref{fig:convnonrotating}, the density profiles from the GPE approach exhibit excellent agreement with those derived from the fluid approach, with deviations well within the numerical uncertainties. This consistency confirms that our wave-based method accurately reproduces the stable density configuration of the non-rotating self-gravitating BEC.

\subsubsection{Rotating Halos}
In the rotating case, we imposed an artificial spherical infinite wall as a hard-wall boundary condition to confine the condensate and stabilize the simulation. The radius of the spherical wall was varied to assess the sensitivity of the density profile and vortex structure to the boundary conditions. We observed negligible changes in both the density profile and the spatial locations of vortices, even for significant variations in the wall radius. These results further demonstrate numerical convergence with respect to the domain size.

\end{document}